\documentclass[aps,pra,showpacs,twocolumn,groupedaddress,floatfix]{revtex4-2}
\usepackage{amsfonts}
\usepackage{amssymb}
\usepackage{amsmath}
\usepackage{graphicx}
\usepackage{epsfig}
\usepackage{color}
\usepackage{xcolor}
\usepackage{tikz}
\usepackage{bm}
\usepackage{booktabs}
\usepackage[colorlinks=true,linkcolor=blue,citecolor=blue,urlcolor=blue]{hyperref}

\definecolor{orcidlogocol}{HTML}{A6CE39}
\DeclareRobustCommand{\orcidicon}[1]{\href{https://orcid.org/#1}{\tikz[baseline=-0.6ex]\node[fill=orcidlogocol,circle,inner sep=0.25pt,minimum size=1.45ex]{\scriptsize\textcolor{white}{\sffamily iD}};}}
\input{tcilatex}
\begin{document}

\title{Generalized $\eta $-pairing eigenstates in three-component Hubbard
models under a transverse field}
\author{F. X. Liu\,\orcidicon{0009-0008-9468-8972}}
\author{Z. Song\,\orcidicon{0000-0002-3315-4589}}
\email{songtc@nankai.edu.cn}
\affiliation{School of Physics, Nankai University, Tianjin 300071, China}

\begin{abstract}
We investigate three-component Hubbard models on a bipartite lattice in the
presence of a transverse field. Unlike the two-component Hubbard model,
where the $\eta$-pairing symmetry survives the transverse field, the
three-component model possesses neither $\eta$-pairing symmetry nor
conservation of the particle number of each component. We introduce a
generalized $\eta$-pairing operator, formed as a hybridization of three
types of two-component pairing operators, and employ the restricted
spectrum-generating algebra to construct a family of exact eigenstates.
These eigenstates exhibit off-diagonal long-range order. Furthermore, we
construct an exact tensor-product state whose coherent periodic dynamics
provide an exact realization of quantum many-body scar dynamics while
simultaneously exhibiting long-range magnetic order. We also propose three
protocols for preparing this state via quench dynamics, and verify their
effectiveness through numerical simulations.
\end{abstract}

\maketitle

\section{Introduction}

The Hubbard model is one of the paradigmatic models in condensed matter
physics, providing a unified framework for studying strongly correlated
electrons, magnetism, superconductivity, and quantum phase transitions \cite%
{Hubbard1963,Essler2005}. Among its remarkable properties is the existence
of the $\eta $-pairing symmetry discovered by Yang \cite{Yang1989}, which
gives rise to a family of exact eigenstates with off-diagonal long-range
order (ODLRO). These $\eta $-pairing states have been extensively studied in
connection with unconventional superconductivity, nonequilibrium dynamics,
and quantum many-body scars \cite%
{Shen1990,Montorsi1992,HDK2025,Moudgalya2022,LFX2026}.

Recent advances in ultracold atomic gases have considerably broadened the
scope of Hubbard physics. Optical lattice experiments have realized
multicomponent fermionic systems with SU($N$) symmetry, enabling the
exploration of strongly correlated phenomena beyond the conventional
two-component Hubbard model \cite{Gorshkov2010,Taie2012,Cazalilla2014,
Hofrichter2016}. Three-component Hubbard models provide the simplest
extension, supporting rich pairing structures, trionic correlations, and
novel quantum phases that have no counterpart in two-component systems \cite%
{Capponi2016,Rapp2007}. Beyond conventional Dirac and Weyl fermions,
crystalline symmetries allow for unconventional quasiparticles such as
triply degenerate fermions and multifold fermions
\cite{Bradlyn2016,Weng2016,Zhu2016,Chang2017}.

In the conventional two-component Hubbard model on a bipartite lattice, the $%
\eta$-pairing symmetry survives the presence of a transverse field that
mixes the two spin components. Consequently, the exact $\eta$-pairing
eigenstates remain valid eigenstates of the Hamiltonian. For three-component
systems, however, a transverse field hybridizes the three internal
components and simultaneously breaks both the conventional $\eta$-pairing
symmetry and the conservation of the particle number of each component.
Whether exact paired eigenstates still exist under these conditions remains
an open question.

Meanwhile, considerable attention has recently been devoted to quantum
many-body scars (QMBS), which are atypical nonthermal eigenstates embedded
in otherwise thermal spectra and exhibit long-lived coherent dynamics \cite%
{Turner2018,Serbyn2021,Moudgalya2022}. Exact scar eigenstates have been
identified in a variety of constrained spin and fermionic models through
mechanisms such as spectrum-generating algebras (SGA) and restricted
spectrum-generating algebras (RSGA) \cite{Mark2020,Choi2019,Moudgalya2020}.
These algebraic approaches have provided a unified framework for
constructing exact eigenstates with sub-volume-law entanglement and
persistent revivals.

In this work, we show that exact paired eigenstates continue to exist in
three-component Hubbard models subjected to a transverse field despite the
absence of the conventional $\eta $-pairing symmetry. We introduce a
generalized $\eta $-pairing operator constructed from a hybridization of
three types of two-component pairing operators. Based on the restricted
spectrum-generating algebra, we obtain three families of exact eigenstates
forming equally spaced energy towers. These states exhibit off-diagonal
long-range order. Furthermore, we identify an exact tensor-product state
that undergoes perfect periodic dynamics and simultaneously possesses
long-range magnetic order, providing an exact realization of quantum
many-body scar dynamics. Finally, we propose three quench protocols for
preparing this state and verify their effectiveness through numerical
simulations.

\section{Model}

\label{sec_model}

We consider a three-component Hubbard model on a bipartite lattice with
sublattices A and B, described by the Hamiltonian%
\begin{equation}
H=\sum_{j\in A\cup B}V_{j}+\sum_{i\in A,j\in B}T_{ij}+\sum_{j\in A\cup
B}\Sigma _{j}^{x},
\end{equation}%
on a bipartite lattice, A and B, which contains three parts classified by
their respective actions. The term%
\begin{equation}
V_{j}=U\sum_{\alpha <\beta }n_{j,\alpha }n_{j,\beta }+\mu \sum_{\alpha
=1}^{3}n_{j,\alpha },  \label{V_j}
\end{equation}%
represents an on-site density-density interaction between $j$th-site
fermions in components $\alpha $ and $\beta $ with strength $U$, and
chemical potentials with strength $\mu $. The term%
\begin{equation}
T_{ij}=-t\sum_{\alpha =1}^{3}c_{i,\alpha }^{\dag }c_{j,\alpha }+\mathrm{H.}%
\text{\textrm{c}}\mathrm{.}.  \label{T_ij}
\end{equation}%
describes fermions hopping between sites $i$ and $j$ with a tunneling
amplitude $t$. The term%
\begin{equation}
\Sigma _{j}^{x}=g\sum_{\alpha <\beta }\left( c_{j,\alpha }^{\dag }c_{j,\beta
}+\mathrm{H.}\text{\textrm{c}}\mathrm{.}\right) ,  \label{Sigma_x}
\end{equation}%
represents transitions between two different components of the $j$th-site
fermion. For simplicity, we refer $\Sigma _{j}^{x}$ to as the transverse
field term, since it contains the sum of three types of pseudo-spin
operators in the $x$ direction. However, we emphasize that it is not, by
itself, a pseudo-spin operator as one imagines at first glance. We will give
an explanation later. Here $c_{j,\alpha }^{\dag }$ and $c_{j,\alpha }$
denote the fermionic creation and annihilation operators of the $\alpha $th
component at site $j$, and $n_{j,\alpha }=c_{j,\alpha }^{\dag }c_{j,\alpha }$
is the number operator. The fermion operators obey the anticommutation
relations $\{c_{i,\alpha },c_{j,\beta }^{\dag }\}=\delta _{i,j}\delta
_{\alpha ,\beta }$ and $\{c_{i,\alpha },c_{j,\beta }\}=0$. The fermions hop
between nearest-neighbor sites $i$ and $j$ with a tunneling amplitude $t$.
The Hamiltonian respects the conservation of the particle number, i.e., 
\begin{equation}
\left[ H,\sum_{j\in A\cup B}\sum_{\alpha =1}^{3}n_{j,\alpha }\right] =0.
\end{equation}%
This allows us to consider the construction of eigenstates in each sector
with a fixed total particle number.

In the absence of the term $\Sigma _{j}^{x}$, the Hamiltonian respects the
conservation of the $\alpha $th component particle number, i.e., $\left[
H,\sum_{j\in A\cup B}n_{j,\alpha }\right] =0$. A set of eigenstates can be
constructed in the invariant subspace with fixed $\alpha $th component
particle number. However, these states are no longer the eigenstates for
nonzero $g$.

\section{$\protect\eta $-pairing eigenstates}

\label{eta-pairing eigenstates}

To obtain the exact eigenstates of the Hamiltonian $H$, we introduce three
sets of generalized $\eta $-pairing operators with $\alpha =1,2,3$, defined
as%
\begin{eqnarray}
\eta _{\alpha }^{+} &=&\left( \eta _{\alpha }^{-}\right) ^{\dag
}=\sum_{j}\eta _{j,\alpha }^{+}, \\
\eta _{\alpha }^{z} &=&\sum_{j}\eta _{j,\alpha }^{z}=\frac{1}{2}\left[ \eta
_{\alpha }^{+},\eta _{\alpha }^{-}\right] .
\end{eqnarray}%
Explicitly, these operators take the form%
\begin{equation}
\left( 
\begin{array}{c}
\eta _{j,1}^{+} \\ 
\eta _{j,2}^{+} \\ 
\eta _{j,3}^{+}%
\end{array}%
\right) =\lambda _{j}\mathcal{V}\left( 
\begin{array}{c}
c_{j,1}^{\dagger }c_{j,2}^{\dagger } \\ 
c_{j,2}^{\dagger }c_{j,3}^{\dagger } \\ 
c_{j,3}^{\dagger }c_{j,1}^{\dagger }%
\end{array}%
\right)  \label{eta-pairing operators}
\end{equation}%
with $\lambda _{j}=1$ for $j\in A$ and $\lambda _{j}=-1$ for $j\in B$. In
the basis $\left( c_{j,1}^{\dagger }c_{j,2}^{\dagger },c_{j,2}^{\dagger
}c_{j,3}^{\dagger },c_{j,3}^{\dagger }c_{j,1}^{\dagger }\right) ^{T}$, the
matrix representation of operator $\Sigma _{j}^{x}$ is%
\begin{equation}
\left[ \Sigma _{j}^{x}\right] =g\left( 
\begin{array}{ccc}
0 & -1 & -1 \\ 
-1 & 0 & -1 \\ 
-1 & -1 & 0%
\end{array}%
\right) .
\end{equation}%
The above matrix is%
\begin{equation}
\mathcal{V}=\left( 
\begin{array}{ccc}
\frac{1}{\sqrt{2}} & -\frac{1}{\sqrt{2}} & 0 \\ 
\frac{1}{\sqrt{6}} & \frac{1}{\sqrt{6}} & -\frac{2}{\sqrt{6}} \\ 
\frac{1}{\sqrt{3}} & \frac{1}{\sqrt{3}} & \frac{1}{\sqrt{3}}%
\end{array}%
\right)
\end{equation}%
which diagonalizes the matrix representation of operator $\Sigma _{j}^{x}$
in the form%
\begin{equation}
\mathcal{V}\left[ \Sigma _{j}^{x}\right] \mathcal{V}^{-1}=g\left( 
\begin{array}{ccc}
1 &  &  \\ 
& 1 &  \\ 
&  & -2%
\end{array}%
\right) .
\end{equation}%
Not surprisingly, $\eta _{\alpha }^{\pm }$\ and $\eta _{\alpha }^{z}$\
satisfy the Lie algebra commutation relations

\begin{equation}
\left[ \eta _{\alpha }^{+},\eta _{\alpha }^{-}\right] =2\eta _{\alpha }^{z},%
\left[ \eta _{\alpha }^{z},\eta _{\alpha }^{\pm }\right] =\pm \eta _{\alpha
}^{\pm }.
\end{equation}%
However, their summation operators $\sum_{\alpha }\eta _{\alpha }^{\pm }$\
and $\sum_{\alpha }\eta _{\alpha }^{z}$\ do not satisfy such commutation
relations, since the operators for different $\alpha $ are not independent.
When the model reduces to a two-component Hubbard model upon eliminating one
of the three $\alpha $\ components, the corresponding $\eta $-pairing
operators read $\eta _{j,1}^{+}=\lambda _{j}c_{j,1}^{\dagger
}c_{j,2}^{\dagger }$, $\eta _{j,2}^{+}=\lambda _{j}c_{j,2}^{\dagger
}c_{j,3}^{\dagger }$, and $\eta _{j,3}^{+}=\lambda _{j}c_{j,3}^{\dagger
}c_{j,1}^{\dagger }$, respectively. Accordingly, the term $\Sigma _{j}^{x}\ $%
is $g\lambda _{j}\sum_{\alpha \neq \beta }\left( \eta _{j,\alpha }^{+}+\eta
_{j,\alpha }^{-}\right) $. Note that the three sets of operators are not
mutually independent. In this sense, term $\Sigma _{j}^{x}$\ is called the
transverse field term for simplicity.

On the other hand, in the absence of the $\Sigma _{j}^{x}$\ term, it was
shown in Ref.~\cite{Nakagawa2024} that the
operators $\sum_{j}\lambda _{j}c_{j,1}^{\dagger }c_{j,2}^{\dagger }$, $%
\sum_{j}\lambda _{j}c_{j,2}^{\dagger }c_{j,3}^{\dagger }$, and $%
\sum_{j}\lambda _{j}c_{j,3}^{\dagger }c_{j,1}^{\dagger }$ can be used to
generate eigenstates of the Hamiltonian. However, once the $\Sigma _{j}^{x}$%
\ term is included, these states cease to be eigenstates. We now show that
the generalized $\eta $ operators, defined in Eq. (\ref{eta-pairing
operators}) generate the exact eigenstates instead.

The proof is based on the restricted spectrum-generating algebra (RSGA). We
first derive the commutation relations between the generalized $\eta $
operators and each term of the Hamiltonian. Diagonalizing the matrix $\left[
\Sigma _{j}^{x}\right] $ yieds the local commutation relation 
\begin{equation}
\left[ \Sigma _{j}^{x},\eta _{j,\alpha }^{+}\right] =g\kappa _{\alpha }\eta
_{j,\alpha }^{+},
\end{equation}%
from which one immediately obtains%
\begin{equation}
\left[ \sum_{j}\Sigma _{j}^{x},\eta _{\alpha }^{+}\right] =g\kappa _{\alpha
}\eta _{\alpha }^{+},  \label{sigma_eta_comm}
\end{equation}%
where $(\kappa _{1},\kappa _{2},\kappa _{3})=(1,1,-2)$ are the eigenvalues
of the matrix $\left[ \Sigma _{j}^{x}\right] $. The commutation relation
between the generalized $\eta $ operators and the kinetic term of the
Hamiltonian is given by%
\begin{equation}
\left[ \sum_{i\in A,j\in B}T_{ij},\eta _{\alpha }^{+}\right] =0.
\label{T_eta_comm}
\end{equation}%
Indeed, for every pair of sites $(i,j)$ on the bipartite lattice, we have $%
\lambda _{i}=-\lambda _{j}$, and the hopping process conserves the component
index. Consequently, the contributions from $\eta _{i,\alpha }^{+}$ and $%
\eta _{j,\alpha }^{+}$ exactly cancel. For the on-site interaction term,
introducing 
\begin{equation}
N_{j}=\sum_{\alpha =1}^{3}n_{j,\alpha },
\end{equation}%
we express the term $V_{j}$\ as%
\begin{equation}
V_{j}=\frac{U}{2}N_{j}\left( N_{j}-1\right) +\mu N_{j},
\end{equation}%
based on the identity%
\begin{equation}
\sum_{\alpha <\beta }n_{j,\alpha }n_{j,\beta }=\frac{1}{2}N_{j}\left(
N_{j}-1\right) .
\end{equation}%
The corresponding commutator is then%
\begin{equation}
\left[ \sum_{j}V_{j},\eta _{\alpha }^{+}\right] =\left( U+2\mu \right) \eta
_{\alpha }^{+}+2U\sum_{j}\eta _{j,\alpha }^{+}N_{j}.
\end{equation}%
Combining the above results, we obtain 
\begin{equation}
\left[ H,\eta _{\alpha }^{+}\right] =\Delta _{\alpha }\eta _{\alpha
}^{+}+2UR_{\alpha }^{+},  \label{RSGA_comm}
\end{equation}%
with 
\begin{equation}
R_{\alpha }^{+}=\sum_{j}\eta _{j,\alpha }^{+}N_{j},
\end{equation}%
and%
\begin{equation}
\Delta _{\alpha }=U+2\mu +g\kappa _{\alpha }.
\end{equation}%
The second term on the right-hand side of Eq. (\ref{RSGA_comm}) vanishes
when acting on the vacuum state, since $R_{\alpha }^{+}\left\vert
0\right\rangle =0$, and therefore we have 
\begin{equation}
\left[ H,\eta _{\alpha }^{+}\right] \left\vert 0\right\rangle =\Delta
_{\alpha }\eta _{\alpha }^{+}\left\vert 0\right\rangle .
\label{restricted_comm}
\end{equation}%
Furthermore, since $\left( \eta _{j,\alpha }^{+}\right) ^{2}=0$ and local
pair operators on different sites commute, we have 
\begin{equation}
\left[ R_{\alpha }^{+},\eta _{\alpha }^{+}\right] =0,
\end{equation}%
which establishes the RSGA condition%
\begin{equation}
\left[ \left[ H,\eta _{\alpha }^{+}\right] ,\eta _{\alpha }^{+}\right] =0.
\label{RSGA_double_comm}
\end{equation}%
Since the vacuum satisfies $H\left\vert 0\right\rangle =0$, Eqs. (\ref%
{restricted_comm}) and (\ref{RSGA_double_comm}) imply 
\begin{equation}
H\left( \eta _{\alpha }^{+}\right) ^{n}\left\vert 0\right\rangle =n\Delta
_{\alpha }\left( \eta _{\alpha }^{+}\right) ^{n}\left\vert 0\right\rangle .
\end{equation}%
The normalized eigenstates are thus

\begin{equation}
\left\vert \psi _{n,\alpha }\right\rangle =\frac{1}{n!\sqrt{C_{N}^{n}}}%
\left( \eta _{\alpha }^{+}\right) ^{n}\left\vert 0\right\rangle ,
\end{equation}%
with $n=0,1,2,...,N$, where $N$ is the total number of sites in the
bipartite lattice.

We note that, for $\mu =-\left( U+g\right) /2$\ $(\alpha =1,2)$ and $\mu
=-\left( U-2g\right) /2$ $\left( \alpha =3\right) $, the corresponding
eigenstates $\left\{ \left\vert \psi _{n,\alpha }\right\rangle \right\} $\
become degenerate at zero energy. These parameter values are referred to as
the resonant conditions. Away from these resonant conditions, the
eigenstates form three equally spaced energy towers of quantum scar states.
In this work, we focus on one of these towers, which exhibits particularly
interesting properties.

We consider the following superposition of the degenerate eigenstates%
\begin{equation}
\left\vert \phi _{\alpha }(\theta )\right\rangle =\sum_{n}d_{n}\left\vert
\psi _{n,\alpha }\right\rangle ,
\end{equation}%
where $d_{n}=\sqrt{C_{N}^{n}}(-i)^{n}\sin ^{n}(\frac{\theta }{2})\cos ^{N-n}(%
\frac{\theta }{2})$. Using the binomial expansion and the commutativity of
pair operators on different sites, one readily obtains%
\begin{equation}
\left\vert \phi _{\alpha }(\theta )\right\rangle
=\dprod\limits_{j=1}^{N}\left\vert \phi _{j,\alpha }(\theta )\right\rangle ,
\end{equation}%
where the local state is given by%
\begin{equation}
\left\vert \phi _{j,\alpha }(\theta )\right\rangle =\left[ \cos (\frac{%
\theta }{2})-i\sin (\frac{\theta }{2})\eta _{j,\alpha }^{+}\right]
\left\vert 0\right\rangle _{j}.
\end{equation}%
It is a tensor product state over all lattice sites. Here, $\theta $ is an
arbitrary angle that parametrizes the state. When the chemical potential
does not satisfy the resonant conditions, the state $\left\vert \phi
_{\alpha }(\theta )\right\rangle $\ is no longer an eigenstate. However, its
time evolution is given by $\left\vert \phi _{\alpha }(\theta
,t)\right\rangle =\dprod\limits_{j=1}^{N}\left\vert \phi _{j,\alpha }(\theta
,t)\right\rangle $, where%
\begin{equation}
\left\vert \phi _{j,\alpha }(\theta ,t)\right\rangle =\left[ \cos (\frac{%
\theta }{2})-ie^{-i\Delta _{\alpha }t}\sin (\frac{\theta }{2})\eta
_{j,\alpha }^{+}\right] \left\vert 0\right\rangle _{j}.
\end{equation}%
The time-evolved state therefore remains a product state and evolves
periodically in time, exhibiting the characteristic long-lived coherent
dynamics of quantum many-body scars. In this work, we focus on the state 
\begin{equation}
\left\vert \phi _{\alpha }(\pi /2)\right\rangle =\frac{1}{2^{N/2}}%
\dprod\limits_{j=1}^{N}(1-i\eta _{j,\alpha }^{+})\left\vert 0\right\rangle ,
\end{equation}%
because it exhibits maximal fluctuations of the local pair number, strong
ODLRO, and long-range magnetic order.

\section{State preparation via quantum quench dynamics}

\label{Dynamic generations}

In this section, we investigate how to prepare the state $|\phi _{\alpha
}(\pi /2)\rangle $ using experimentally feasible protocols. We propose three
different quench schemes to generate this state through nonequilibrium
dynamics, which constitutes a crucial step toward a coherent experimental
realization of the proposed quantum states \cite%
{Polkovnikov2011,Greiner2002,Bloch2008}.

The basic idea is to initialize the system in an easily prepared eigenstate
of a prequench Hamiltonian $H_{0}$ and subsequently perform a quantum quench
from $H_{0}$ to the postquench Hamiltonian $H_{\mathrm{pq}}$. Here, $H_{0}$
is obtained from the original Hamiltonian $H$ by tuning specific model
parameters. In the present work, we choose the chemical potential $\mu $
such that all states in the degenerate manifold $\{|\psi _{n,\alpha }\rangle
\}$ become zero-energy eigenstates of $H_{0}$. Consequently, the quench
dynamics provides a controlled route to access the coherent superposition of
degenerate eigenstates, similar to the experimental preparation protocols
used for quantum many-body scars and other nonthermal quantum states.

After the quench, the Hamiltonian becomes

\begin{equation}
H_{\mathrm{pq}}=H_{0}+H_{I}.  \label{H_pq}
\end{equation}%
In the following, we consider three different forms of the driving term $%
H_{I}$ for preparing the state $|\phi _{\alpha }(\pi /2)\rangle $,

\begin{equation}
H_{I}=\sum_{j=1}^{N}\mathbf{B}_{j}\cdot \boldsymbol{\eta }_{j,\alpha },
\end{equation}%
where $\boldsymbol{\eta }_{j,\alpha }=(\eta _{j,\alpha }^{x},\eta _{j,\alpha
}^{y},\eta _{j,\alpha }^{z})$ denotes the pseudospin operator of type $%
\alpha $, and $\mathbf{B}_{j}$ represents an on-site driving field, which
can be either Hermitian or non-Hermitian. The quench dynamics provides a
versatile approach for preparing nontrivial many-body states and exploring
nonequilibrium quantum phenomena~\cite{Polkovnikov2011}. Moreover,
non-Hermitian driving has recently emerged as a powerful tool for
controlling quantum states and engineering novel quantum phases \cite%
{Ashida2020,Bergholtz2021}.

Specifically, we investigate three protocols: (i) a Hermitian driving term,
(ii) a globally non-Hermitian driving term, and (iii) a locally
non-Hermitian driving term. Each protocol provides a distinct route for
generating the target state, while exhibiting its own advantages and
limitations in terms of experimental accessibility and state-preparation
efficiency.

\subsection{Hermitian system}

\label{Hermitian system}

To prepare the state $|\phi_{\alpha}(\pi/2)\rangle$, we first consider a
Hermitian postquench Hamiltonian characterized by

\begin{equation}
\mathbf{B}_{j}=(-1)^{j}B_{0}(1,0,0),
\end{equation}%
where $B_{0}$ is a constant driving strength. The interaction part of the
postquench Hamiltonian is then given by

\begin{equation}
H_{I}=\frac{B_{0}}{2}(\eta _{\alpha }^{+}+\eta _{\alpha }^{-}).
\end{equation}%
Using the SU(2) algebra of the pseudospin operators, one obtains

\begin{eqnarray}
H_{I}|\psi _{n,\alpha }\rangle &=&\frac{B_{0}}{2}[\sqrt{(n+1)(N-n)}|\psi
_{n+1,\alpha }\rangle  \notag \\
&&+\sqrt{n(N-n+1)}|\psi _{n-1,\alpha }\rangle ],
\end{eqnarray}%
for $n=0,1,\cdots ,N$, with the boundary convention

\begin{equation}
|\psi _{-1,\alpha }\rangle =|\psi _{N+1,\alpha }\rangle =0.
\end{equation}%
This relation demonstrates that the subspace spanned by $\{|\psi _{n,\alpha
}\rangle \}$ is invariant under the action of $H_{I}$. Therefore, the
dynamics generated by $H_{I}$ is restricted to this $(N+1)$-dimensional
subspace.

In this basis, the matrix representation of $H_I$ has only off-diagonal
nonzero elements,

\begin{eqnarray}
(M)_{N+1-n,N-n} &=&(M)_{N-n,N+1-n}  \notag \\
&=&\frac{B_{0}}{2}\sqrt{(n+1)(N-n)},
\end{eqnarray}%
where $n=0,1,\cdots ,N-1$. Starting from the initial state

\begin{equation}
|\Phi (0)\rangle =|\psi _{0,\alpha }\rangle \equiv |0\rangle ,
\end{equation}%
the time evolution remains confined within this invariant subspace. More
importantly, the evolved state can be exactly factorized as

\begin{equation}
|\Phi _{\alpha }(t)\rangle =\prod_{j}|\Phi _{j,\alpha }(t)\rangle ,
\end{equation}%
where the local state is

\begin{equation}
|\Phi_{j,\alpha}(t)\rangle = \left[ \cos\left(\frac{B_0t}{2}\right)
-i\sin\left(\frac{B_0t}{2}\right) \eta_{j,\alpha}^{+} \right] |0\rangle_j .
\end{equation}

Several important properties follow from this exact evolution. (i) The state 
$|\Phi_{\alpha}(t)\rangle$ remains an eigenstate of $H_0$ because all states
in the manifold $\{|\psi_{n,\alpha}\rangle\}$ are degenerate zero-energy
eigenstates. (ii) The evolution can be interpreted as a collective
pseudospin rotation generated by the decoupled on-site Hamiltonian
corresponding to $T_{ij}=0$. (iii) At special evolution times

\begin{equation}
t=\frac{(4m+1)\pi }{2B_{0}},
\end{equation}%
the state exactly reaches the target state $|\phi _{\alpha }(\pi /2)\rangle $%
.

The underlying mechanism can be understood from the annihilation property of
the hopping term. For any nearest-neighbor pair $(i,j)$, one finds

\begin{align}
& T_{ij}|\Phi _{i,\alpha }(t)\rangle |\Phi _{j,\alpha }(t)\rangle  \notag \\
=& T_{ij}\left[ \cos ^{2}\left( \frac{B_{0}t}{2}\right) -\sin ^{2}\left( 
\frac{B_{0}t}{2}\right) \eta _{i,\alpha }^{+}\eta _{j,\alpha }^{+}\right]
|0\rangle _{i}|0\rangle _{j}  \notag \\
& -\frac{i}{2}\sin (B_{0}t)T_{ij}(\eta _{i,\alpha }^{+}+\eta _{j,\alpha
}^{+})|0\rangle _{i}|0\rangle _{j}  \notag \\
=& 0.
\end{align}%
Therefore, the hopping processes are completely suppressed in this
dynamically generated state, ensuring that the product-state structure is
preserved during the evolution.

\subsection{Non-Hermitian system}

\label{Non-Hermitian system}

In general, the real-time dynamics of Hermitian many-body quantum systems is
difficult to control due to the complexity generated by particle-particle
interactions. In particular, the evolved states are generally not
analytically predictable. This difficulty can be circumvented in properly
designed non-Hermitian systems, where exceptional points (EPs) induce
nontrivial dynamical behavior: different eigenstates coalesce into a single
defective eigenstate, which can dominate the long-time evolution under
appropriate conditions~\cite%
{Berry2004,Heiss2012,Miri2019,Zhang2020,Wang2021,YXM2019,YXM2022}.

Specifically, to prepare the state $|\phi _{\alpha }(\pi /2)\rangle $, we
consider a non-Hermitian postquench Hamiltonian characterized by

\begin{equation}
\mathbf{B}_{j}=(-1)^{j}B_{0}(1,0,-i),
\end{equation}%
where $B_{0}$ is a constant driving strength. The interaction part becomes

\begin{equation}
H_{I}=\frac{B_{0}}{2}(\eta _{\alpha }^{+}+\eta _{\alpha }^{-}-2i\eta
_{\alpha }^{z}).
\end{equation}%
Within the invariant subspace spanned by $\{|\psi _{n,\alpha }\rangle \}$,
the matrix representation of $H_{I}$ is an $(N+1)\times (N+1)$ matrix with
nonzero off-diagonal elements

\begin{eqnarray}
(\mathcal{M})_{N+1-n,N-n} &=&(\mathcal{M})_{N-n,N+1-n}  \notag \\
&=&\frac{B_{0}}{2}\sqrt{(n+1)(N-n)},
\end{eqnarray}%
where $n=0,1,\cdots ,N-1$, and diagonal elements

\begin{equation}
(\mathcal{M})_{N+1-n,N+1-n}=\frac{iB_{0}}{2}(2n-N),
\end{equation}%
where $n=0,1,\cdots ,N$.

It is evident that $\mathcal{M}$ is a nilpotent matrix satisfying

\begin{equation}
\mathcal{M}^{N+1}=0,
\end{equation}%
which corresponds to a single $(N+1)$-dimensional Jordan block at the
exceptional point. The time-evolution operator within this subspace is
therefore given by

\begin{equation}
U(t)=e^{-i\mathcal{M}t} = \sum_{l=0}^{N} \frac{1}{l!}(-i\mathcal{M}t)^l .
\end{equation}

Starting from the initial state

\begin{equation}
|\Phi (0)\rangle =|\psi _{0,\alpha }\rangle \equiv |0\rangle ,
\end{equation}%
the dynamics remains restricted to this invariant subspace. The evolved
state can be factorized as

\begin{equation}
|\Phi _{\alpha }(t)\rangle =\prod_{j}|\Phi _{j,\alpha }(t)\rangle ,
\end{equation}%
where the local state is

\begin{equation}
|\Phi_{j,\alpha}(t)\rangle = \left( 1+\frac{B_0t}{2} -i\frac{B_0t}{2}%
\eta_{j,\alpha}^{+} \right)|0\rangle_j .
\end{equation}

Several important properties follow from this evolution. (i) $|\Phi _{\alpha
}(t)\rangle $ remains an eigenstate of $H_{0}$, since all states in the
manifold $\{|\psi _{n,\alpha }\rangle \}$ are degenerate zero-energy
eigenstates. (ii) The dynamics can be interpreted as the evolution generated
by the decoupled on-site Hamiltonian of $H_{0}+H_{I}$ in the limit $T_{ij}=0$%
. (iii) In the long-time limit, the state approaches the coalescent state,

\begin{equation}
|\Phi _{\alpha }(t\rightarrow \infty )\rangle \rightarrow |\phi _{\alpha
}(\pi /2)\rangle .
\end{equation}%
The underlying mechanism is again the annihilation property of the hopping
term,

\begin{equation}
T_{ij}|\Phi _{i,\alpha }(t)\rangle |\Phi _{j,\alpha }(t)\rangle =0.
\end{equation}

\subsection{Single non-Hermitian impurity}

\label{Single non-Hermitian impurity}

The essential difference between the Hermitian and non-Hermitian protocols
is that the former exhibits periodic evolution, whereas the latter generates
a directional evolution toward a coalescent state. Interestingly, the
high-order exceptional point (EP) structure survives even when the
non-Hermitian field is applied only locally. Specifically, we introduce a
complex field at a single site $l$,

\begin{equation}
\mathbf{B}_{j}=B_{0}\delta _{jl}(1,0,-i).  \label{single impurity}
\end{equation}

Note that the vacuum state $|0\rangle$ and the target state $%
|\phi_{\alpha}(\pi/2)\rangle$ are degenerate eigenstates of the Hermitian
Hamiltonian $H_0$. Moreover, we have

\begin{equation}
H_{I}|\phi _{\alpha }(\pi /2)\rangle =0,\qquad H_{I}^{\dagger }|\phi
_{\alpha }(-\pi /2)\rangle =0.
\end{equation}%
Therefore, $|\phi _{\alpha }(\pi /2)\rangle $ and $|\phi _{\alpha }(-\pi
/2)\rangle $ form a pair of biorthogonal conjugate states. Their
biorthogonal overlap satisfies

\begin{equation}
\langle \phi _{\alpha }(-\pi /2)|\phi _{\alpha }(\pi /2)\rangle =0,
\end{equation}%
which indicates the self-orthogonality characteristic of an exceptional
point. Consequently, $|\phi _{\alpha }(\pi /2)\rangle $ and $|\phi _{\alpha
}(-\pi /2)\rangle $ correspond to the coalescent states of $H$ and $%
H^{\dagger }$, respectively.

From the dynamical perspective, in the long-time limit,

\begin{equation}
\begin{split}
\lim_{t\rightarrow \infty }e^{-iHt}|0\rangle & =|\phi _{\alpha }(\pi
/2)\rangle , \\
\lim_{t\rightarrow \infty }e^{-iH^{\dagger }t}|0\rangle & \rightarrow |\phi
_{\alpha }(-\pi /2)\rangle .
\end{split}%
\end{equation}%
However, this dynamical behavior alone does not determine the order of the
exceptional point. Therefore, we further analyze the EP structure using
perturbation theory. Within the degenerate subspace spanned by $\{|\psi
_{n}\rangle \}$, the effective Hamiltonian is represented by an $(N+1)\times
(N+1)$ matrix $\mathcal{H}$. Its nonzero elements are obtained from the
corresponding Jordan-block matrix $\mathcal{M}$,

\begin{equation}
\mathcal{H}=\frac{1}{N}\mathcal{M}.  \label{Perturbation H}
\end{equation}

For numerical verification, we perform exact diagonalization. Under a
non-Hermitian Hamiltonian, the density matrix obeys the generalized von
Neumann equation,

\begin{equation}
i\frac{\partial \rho (t)}{\partial t}=H_{\mathrm{pq}}\rho (t)-\rho (t)H_{%
\mathrm{pq}}^{\dagger }.
\end{equation}%
The formal solution is

\begin{equation}
\rho (t)=e^{-iH_{\mathrm{pq}}t}\rho (0)e^{iH_{\mathrm{pq}}^{\dagger }t}.
\end{equation}%
Since the evolution is non-unitary, the density matrix is normalized as \cite%
{Brody2012,Kawabata2017}

\begin{equation}
\rho (t)=\frac{e^{-iH_{\mathrm{pq}}t}\rho (0)e^{iH_{\mathrm{pq}}^{\dagger }t}%
}{\mathrm{Tr}[e^{-iH_{\mathrm{pq}}t}\rho (0)e^{iH_{\mathrm{pq}}^{\dagger }t}]%
}.
\end{equation}%
We characterize the convergence toward the target state using the Uhlmann
fidelity~\cite{Uhlmann1976,Jozsa1994},

\begin{equation}
F(t)=\left[ \mathrm{Tr}\sqrt{\sqrt{\rho _{\mathrm{h}}}\rho (t)\sqrt{\rho _{%
\mathrm{h}}}}\right] ^{2}.
\end{equation}%
The target state is%
\begin{equation}
\rho _{\mathrm{h}}=\left\vert \phi _{\alpha }(\pi /2)\right\rangle
\left\langle \phi _{\alpha }(\pi /2)\right\vert .
\end{equation}

The long-time behavior of the fidelity can be understood intuitively. In
general, an initial mixed state $\rho(0)$ with equal weights in the
degenerate manifold $\{|\psi_{n,\alpha}\rangle\}$ is expected to evolve
toward the coalescent state under the non-Hermitian dynamics. Therefore,
after proper normalization, the fidelity approaches

\begin{equation}
F(\infty)\approx 1 .
\end{equation}

To verify this prediction, we consider two types of initial states: (i) the
vacuum state,

\begin{equation}
\rho (0)=|0\rangle \langle 0|,  \label{initial 1}
\end{equation}%
and (ii) the maximally mixed pairing state,

\begin{equation}
\rho (0)=\frac{1}{N+1}\sum_{n=0}^{N}|\psi _{n,\alpha }\rangle \langle \psi
_{n,\alpha }|.  \label{initial 2}
\end{equation}

We then calculate the fidelity dynamics $F(t)$ under the full non-Hermitian
Hamiltonians constructed from the single-impurity Hamiltonian in Eq.~(\ref%
{single impurity}) and the perturbative non-Hermitian Hamiltonian in Eq.~(%
\ref{Perturbation H}).

The long-time fidelity can be understood from the spectral property of the
exceptional point. Since all states within the degenerate manifold $%
\{|\psi_{n,\alpha}\rangle\}$ are dynamically attracted to the same
coalescent state, an arbitrary initial state with nonzero projection onto
this manifold evolves toward $|\phi_{\alpha}(\pi/2)\rangle$. Consequently,
after normalization, the asymptotic fidelity satisfies $F(\infty)\approx1$.

\begin{figure*}[tbp]
\centering
\includegraphics[width=0.96\textwidth]{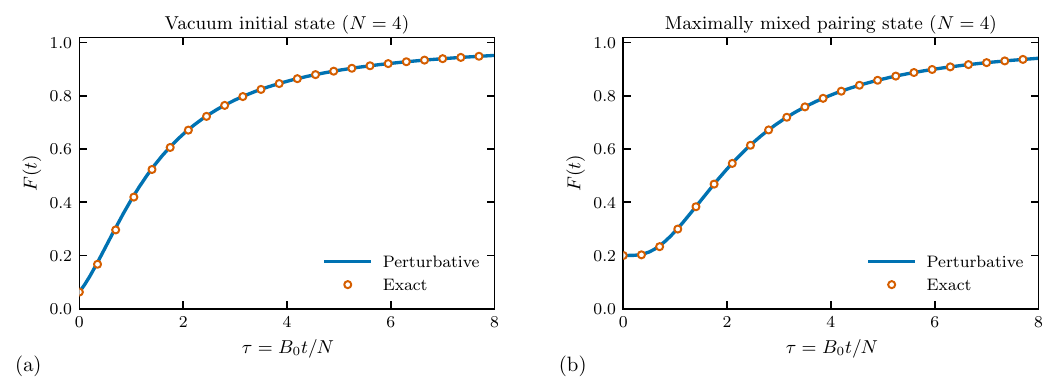}
\caption{Fidelity dynamics $F(t)$ for (a) the vacuum initial state in Eq.~(%
\protect\ref{initial 1}) and (b) the maximally mixed pairing state in Eq.~(%
\protect\ref{initial 2}). The solid curves (Perturbative) show the
first-order projected dynamics obtained from Eq.~(\protect\ref{Perturbation
H}), whereas the open circles (Exact) show the exact time evolution in the
full even-parity Fock space under the single-impurity Hamiltonian in Eq.~(%
\protect\ref{single impurity}). The system size is $N=4$, and the
non-Hermitian impurity field is $\mathbf{B}_{j}=0.1\,\protect\delta%
_{j0}(1,0,-i)$. In both panels, the fidelity approaches unity at long times,
demonstrating the robustness of the EP-induced state preparation.}
\label{fig3}
\end{figure*}

The value of $F\left( t\right) $ after a sufficient long time can be
estimated intuitively. In general, an initial mixed state $\rho \left(
0\right) $ contains equal-amplitude components in each state of $\left\{
\left\vert \psi _{n,\alpha }\right\rangle \right\} $. Then we always have $%
F\left( \infty \right) \approx 1$. The fidelity curves in Fig.~\ref{fig3}
demonstrate the dynamical convergence induced by the single non-Hermitian
impurity. The evolved states from both initial conditions rapidly approach
the target state after sufficiently long evolution time. Importantly,
neither the initial states nor the selected non-Hermitian impurity requires
prior knowledge of the prequench Hamiltonian $H_{0}$. The results
also demonstrate that the perturbation method provides an efficient approach
for investigating the dynamics near exceptional points.

\section{Summary}

\label{Summary}

In summary, we have investigated three-component Hubbard models on a
bipartite lattice in the presence of a transverse field. In contrast to the
two-component Hubbard model, where the conventional $\eta $-pairing symmetry
survives the transverse field, the three-component model possesses neither
the $\eta $-pairing symmetry nor conservation of the particle number of each
component. We have shown that these symmetry-breaking effects do not
preclude the existence of exact paired eigenstates.

By introducing a generalized $\eta$-pairing operator as a hybridization of
three types of two-component pairing operators, we established a restricted
spectrum-generating algebra (RSGA) for the model. Based on this algebraic
structure, we constructed three families of exact eigenstates forming
equally spaced energy towers. Under resonant conditions, these eigenstates
become degenerate zero-energy states, while away from resonance they
constitute exact quantum scar towers. We further showed that these
eigenstates possess off-diagonal long-range order, extending the concept of $%
\eta$ pairing beyond the conventional two-component Hubbard model.

Furthermore, we constructed an exact tensor-product state as a coherent
superposition of the generalized $\eta$-pairing eigenstates. Away from the
resonant conditions, this state exhibits perfect periodic dynamics while
remaining a product state throughout the evolution, providing an exact
realization of quantum many-body scar dynamics. We also demonstrated that a
particular member of this family simultaneously exhibits maximal local
pair-number fluctuations, strong off-diagonal long-range order, and
long-range magnetic order.

Finally, we proposed three quench protocols for preparing the tensor-product
scar state and verified their effectiveness by numerical simulations. Our
results establish a new framework for generalizing $\eta$ pairing to
multicomponent Hubbard systems without conventional pairing symmetry, and
provide an exactly solvable platform for exploring the interplay among
generalized pairing, quantum many-body scars, nonequilibrium coherent
dynamics, and long-range order.

\bibliography{references}

\end{document}